\documentclass[graybox,natbib,nosecnum]{svmult}
\bibpunct{(}{)}{;}{a}{}{,} % suppress commas between author-names and year

\pdfoutput=1   %forces use of pdflatex. Disable if you prefer to use .eps or .ps figures.
\usepackage{mathptmx}       % selects Times Roman as basic font
\usepackage{helvet}         % selects Helvetica as sans-serif font
\usepackage{courier}        % selects Courier as typewriter font
\usepackage{type1cm}        % activate if the above 3 fonts are
\usepackage{makeidx}         % allows index generation
\usepackage{graphicx}        % standard LaTeX graphics tool
\usepackage{units}        %
\usepackage{amssymb}        %
\usepackage{multicol}        % used for the two-column index
\usepackage[bottom]{footmisc}% places footnotes at page bottom
\usepackage[normalem]{ulem}	% for strike-through of text with \sout{}  
\usepackage{hyperref}  %for hyperlinks

\usepackage{soul}   % for high-lighting of text
\usepackage{comment}
\newcommand{\hbindex}[0]{}  % no highlights

\newcommand{\added}[0]{} % no markup of added text

\makeindex             % used for the subject index
\begin{document}

\title*{Transit Timing and Duration Variations for the Discovery and Characterization of Exoplanets \added{in the TESS era}}
\titlerunning{Transit Timing for Characterization and Discovery } 
\author{Eric Agol and Daniel C.\ Fabrycky}
% Use 
\authorrunning{Agol \& Fabrycky} 
\institute{Eric Agol \at Department of Astronomy, Box 351580, University of Washington, Seattle, WA 98195-1580, USA \email{agol@uw.edu}
\and Daniel C.\ Fabrycky \at Dept.\ of Astronomy \& Astrophysics, University of Chicago, Chicago, IL 60637, USA \email{fabrycky@uchicago.edu}}
\maketitle

\abstract{Transiting exoplanets in multi-planet systems have non-Keplerian orbits which can cause the times and durations of transits to vary.  The theory and observations of transit timing variations (TTV) and transit duration variations (TDV) are reviewed.  The \emph{Kepler} spacecraft has detected several hundred perturbed planets, \added{many of which are still undergoing further study, and now TESS is adding to this sample}.  In a few cases, these data have been used to discover additional planets, similar to the historical discovery of Neptune in our own Solar System.  However, the more impactful aspect of TTV and TDV studies has been characterization of planetary systems in which multiple planets transit.  After addressing the equations of motion and parameter scalings, the main dynamical mechanisms for TTV and TDV are described, with citations to the observational literature for real examples.  Constraints on model parameters from timing are elucidated, particularly the origin of the mass/eccentricity degeneracy and how it is overcome by the high-frequency component of the signal.  On the observational side, derivation of timing precision and introduction to the timing diagram are given.  Science results are reviewed, with an emphasis on mass measurements \added{in multi-transiting planetary systems,} from which bulk compositions may be inferred.  \added{The progress being made in studying transit timing with TRAPPIST-1 and TESS multi-planet systems is reviewed, as well as what the future may hold.}}

\section{Introduction}

Transit Timing Variations (\hbindex{TTV}) and Transit Duration Variations (\hbindex{TDV}) are two of the newer tools in the exoplanetary observer's toolbox for discovering and characterizing planetary systems. Like most such tools, they rely on indirect inferences, rather than detecting light from the planet directly.  However, the amount of dynamical information they encode is extremely rich. 

To decode this information, let us start with the dynamical concepts.  Consider the vector stretching from the star of mass $m_0$ to the planet of mass $m$ to be $\mathbf{r}=(x,y,z)$, with a distance $r$ and direction $\mathbf{\hat r}$.  The Keplerian potential per reduced mass, $\phi=-GM/r$ (where $M \equiv m_0 + m$ and the planet is replaced with a body of reduced mass $\mu \equiv m_0 m /M$), gives rise to closed orbits.  This means that, in the absence of perturbations, the trajectory is strictly periodic, $\mathbf{r}(t+P) = \mathbf{r}(t)$.  Moreover, Kepler showed that Tycho Brahe's excellent data for planetary positions were consistent with Copernicus' idea of a heliocentric system only if the planets (including the Earth) followed elliptical paths of semi-major axis $a$, and one focus on the Sun. Newton was successful at finding the principle underlying such orbits, a force law $\mathbf{F} = \mu \mathbf{\ddot r} =-G \mu m_0 r^{-2} \mathbf{\hat r}$, which results in a period $P = 2 \pi a^{3/2} (GM)^{-1/2}$ (i.e. with the $a$-scaling Kepler found the planets actually obeyed).

This research program was thrown into some doubt by the ``Great Inequality,'' the fact that the orbits of Jupiter and Saturn did not fit the fixed Keplerian ellipse model.  This obstacle was overcome by the perturbation theory of Laplace, who used the masses derived via their satellite orbits to explain the deviations of their heliocentric orbits \citep{Wilson1985}.  The insight can be calculated by writing an additional force to that of gravity of the Sun: 
\begin{equation}
\mathbf{F_{1}} = \mu_1 \mathbf{\ddot r_1} = -G \mu_1 M r_{1}^{-2} \mathbf{\hat r_{1}} + \mathbf{F_{12}},
\end{equation}
where now the forces and distances specifically pertain to planet 1, and a force of planet 2 on planet 1 is added.  This latter force consists of two terms: 
\begin{equation}
\mathbf{F_{12}}  = G \mu_1 m_2 \vert r_{2}-r_{1}\vert^{-3} (\mathbf{r_{2}} - \mathbf{r_{1}}) - G \mu_1 m_2 r_{2}^{-2} \mathbf{\hat r_{2}}.
\end{equation}
The first term on the right-hand-side is the direct gravitational acceleration of planet 1 due to planet 2.  The second is an indirect frame-acceleration effect, due to the acceleration the star feels due to the second planet.  Since the Sun is fixed at the zero of the frame, this acceleration is modelled by acceleration of planet 1 in the opposite direction.  \added{This $F_{12}$ force can be used to account for the Great Inequality with the interactions of Jupiter and Saturn.}

Likewise, Leverrier and Adams used planet-planet perturbations in the first discovery of a planet by gravitational means \citep{Adams1847,LeVerrier1877}. In this case, they did not know the zeroth order solution (i.e. the Keplerian ellipse) for the perturber, Neptune.  In its place, they assumed the Titius-Bode rule held, and sought only the phase of the orbit.  This technique worked because they only wanted to see how the acceleration, and then deceleration, of Uranus as it passed Neptune, would betray Neptune's position on the sky to optical observers. The task of discovering planets by TTV is more demanding as one does not have any hints as to what the planet's orbit might be, i.e. one cannot assume it is on a circular orbit or obeys some spacing law. The observation of a single orbit is insufficient for a detection: times of \added{at} least three transits are needed to measure a period change. However, due to measurement error, in only a small fraction of cases is the high-frequency ``chopping'' signal (see Chopping section below) statistically significant after just three transits.  Moreover, the sampling of the orbit only at transit phase causes aliasing of the dynamical signals \citep{Kipping2020a}.

The times of transit are primarily constrained by the decline of stellar flux during transit ingress, and the rise over egress, which occur on a timescale 
\begin{equation} \label{ingress}
\tau \approx \pi^{-1} P(R_p/a) \approx 2.2 {\rm min} \left(\frac{R_p}{R_\oplus}\right) \left(\frac{m_0}{\rm{M_{sun}}}\right)^{-1/3} \left(\frac{P}{10 \rm{d}}\right)^{1/3},
\end{equation}  
assuming a circular orbit, edge-on to the line of sight (impact parameter of $b=0$), around a star of mass $m_0$; usually timing precision can be measured better than this timescale.
This timing precision gives a sensitive measure of the variation of the angular position of a planet relative to a Keplerian orbit.  In contrast, the other dynamical techniques rely on a signal spread through the orbital timescale $P$, and thus the precision of the orbital phase is poorly constrained unless the measurements are of high precision or long duration (although these conditions have been achieved by pulsar timing in PSR 1257 +12 which detected a Great Inequality, \citealt{1994Sci...264..538W}, and by radial velocity in GJ 876 which detected resonant orbital precession, \citealt{Laughlin2001}).

\begin{figure}
\centerline{
\includegraphics[width=0.9\textwidth]{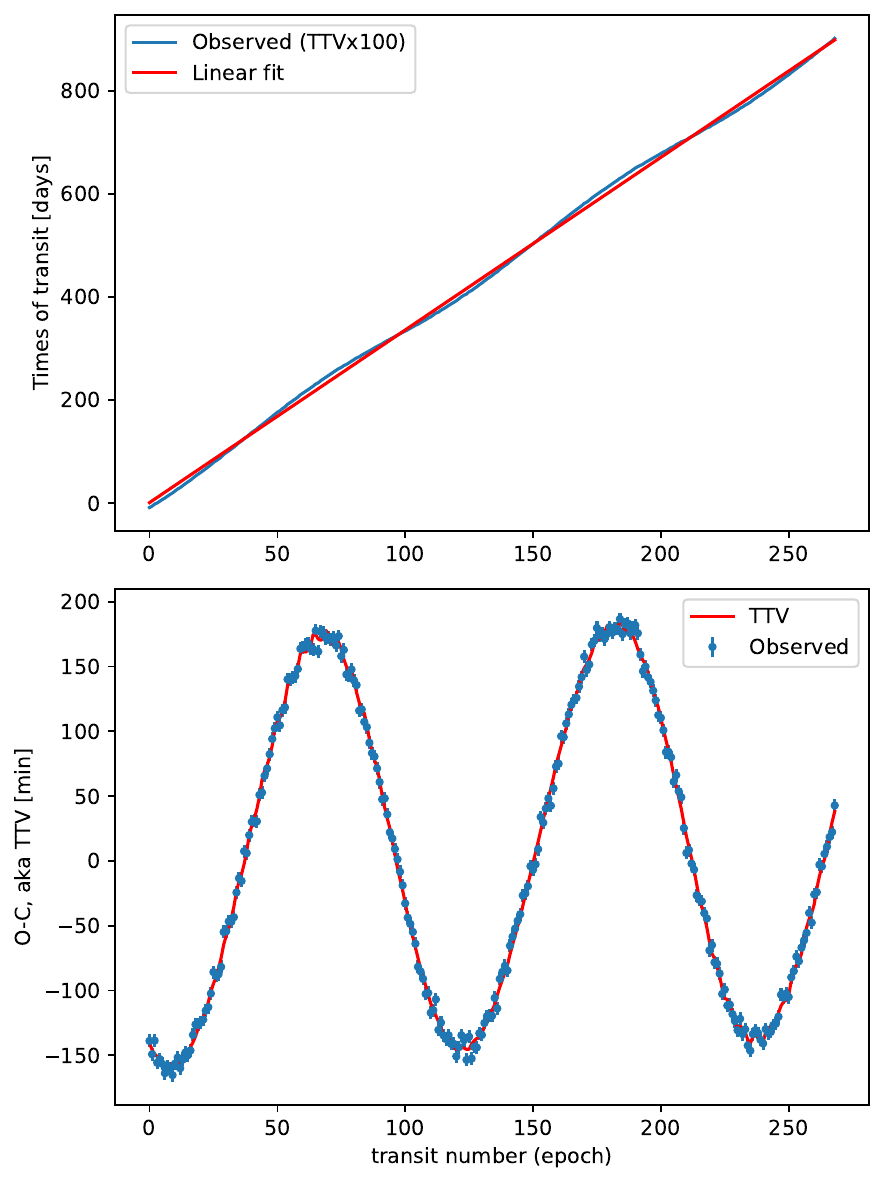}}
\caption{An example of \added{simulated} timing data \added{with added 5-minute normal white noise}.  \emph{Top panel}: the ``measured" midtimes of exoplanet transits \added{(blue; the TTVs have been exaggerated by a factor of 100 in this panel to make them visible by eye)}, to which a line is fit by least-squares \added{(red)}.  \emph{Bottom panel}: the residuals of \added{the linear} fit \added{(blue error bars)}, which is the conventional observed minus calculated ($O-C$) diagram; the original \added{noise-free TTVs are} %\deleted{, to which Gaussian noise was added, is}
also plotted as a \added{red curve}. }
\label{omc}       % Give a unique label
\end{figure}

Orbital positions or transit times are expressed in a table called an \hbindex{ephemeris}. Perturbations cause motions or timing deviations from a Keplerian reference model, especially changes to its instantaneous semimajor axis $a$, eccentricity $e$, \added{time of periastron passage $t_p$}, and longitude of periastron $\omega$.  The latter angle is between the position of closest approach and a plane perpendicular to the line of sight that contains either the primary body or the center of mass.  In the case of transit timing variations, the Keplerian alternative is simply an ephemeris with a constant transit period, $P$:
\begin{equation}
C = T_0 + P \times E, 
\end{equation}
where $E$ is the epoch -- an integer transit number -- and $T_0$ is the time of the transit numbered $E=0$; $C$ stands for ``calculated'' based on a constant-period model.  Meanwhile, the Observed times of transit are denoted $O$.  This notation leads to an \hbindex{$O-C$} (pronounced ``O minus C''; \citealt{Sterken2005}) diagram, in which only the perturbation part is plotted.  An instructive version, modelled after the timing of \added{TOI-2015b} \citep{Jones2024} but with a greatly exaggerated perturbation, is shown in figure~\ref{omc}.  The transit times come earlier than the linear model for transit numbers %\deleted{0-3 and 11-14} 
\added{0-36, 97-151, and 212-263}, and later than the linear model for \added{the remaining} transits. %\deleted{numbers 4-10}.  
These deviations from a constant transit period are what astronomers call TTVs.

The other dynamical effect addressed by this review is TDVs.  Like TTVs, the cause can be changes in $a$, $e$, or $\omega$.  The most dramatic effect, however, is due to orbital plane reorientation.  The angle the orbital plane's normal vector makes to the observer's line of sight --- the inclination, $i$ --- determines the length of the transit chord.  Changes in the inclination will change the length of that chord, which in turn changes the amount of time the planet remains in transit: duration variations. 

The literature on exoplanets has a history of rediscovering effects that had been well studied in the field of binary and multiple stars.  In the current focus, it has long been known to eclipsing-binary observers that long-term depth changes can result from the torque of a third star orbiting the pair \citep{Mayer1971}.  This effect owes to the secular and tidal dynamics which dominate triple star systems \citep{Borkovits2003,Borkovits2011}, dictated by their hierarchical configuration which allows them to remain stable. TDV due to perturbing planets is simply its exoplanetary analogue \citep{2002ApJ...564.1019M}.

The first recognition of the importance of transit timing and duration variations was at the DPS and AAS meetings almost three decades ago by \citet{1996DPS....28.1208D,1996BAAS...28.1112D}, followed a few years later by \citet{2002ApJ...564.1019M} and \citet{Schneider2003,Schneider2004}.  More detailed studies that included the important effect of mean-motion resonance, in which the ratio of two planets' orbital periods is close to the ratio of small integers, were independently investigated by \citet{Holman2005} and \citet{Agol2005}.  The former paper showed that Solar-system like perturbations might be used to find Earth-like planets, should transit times be measured with sufficient accuracy.  The latter paper coined the term \hbindex{`transit-timing variations,'} with acronym TTV, and defined TTVs as the observable accumulation of transit period changes (i.e.\ $O-C$).

Initial studies of TTVs of hot Jupiters were able to place limits on the presence of Earth-mass planets near \hbindex{mean-motion resonance} \citep{2005MNRAS.364L..96S}.  Some further studies claimed detection of perturbing planets causing TTVs or TDVs, but each of these were quickly disputed or refuted by additional measurements.  The first convincing detection awaited the launch of the \emph{Kepler} spacecraft, and the discovery of Kepler-9 which showed large-amplitude TTVs of two Saturn-sized planets with strong significance \citep{2010Sci...330...51H}; this discovery was remarkably similar to predictions based upon the GJ 876 system \citep{Agol2005}.  The Kepler-9 paper kicked off a series of discoveries of TTVs with the \emph{Kepler} spacecraft, with now more than 375 planets displaying TTVs, according to the Confirmed Planets Table in the NASA Exoplanet Archive on 12/5/2024, and some showing TDVs \citep{Holczer2016}.  \added{Although most TTV detections have been space-based, the first ground-based
detections were made for TRAPPIST-1 \citep{Gillon2017} and WASP-148 b \citep{Hebrard2020}.}

\section{Preliminaries}

Since the gravitational interactions between planets occur on the orbital timescale, the
amplitude of TTVs is proportional to the orbital period of each planet,
times a function of other dimensionless quantities.  Thanks to Newton's second law
and Newton's law of gravity, the acceleration of a body does not depend on its own mass.
Thus, the TTVs of each planet scale with the masses of the {\it other} bodies
in the system.
In a two-planet system, then, to lowest order in mass ratio, the $O-C$ formulae are: 
\begin{eqnarray}
\delta t_1 &=& P_1 \frac{m_2}{m_0} f_{12}(\alpha_{12},\mathbf{\theta}_{12}),\cr
\delta t_2 &=& P_2 \frac{m_1}{m_0} f_{21}(\alpha_{12},\mathbf{\theta}_{21}),
\end{eqnarray}
where the masses of the star and planets are $m_0, m_1,$ and $m_2$, and $f_{ij}$ describes the perturbations of planet $j$ on planet $i$,
which is a function of the semi-major axis ratio, $\alpha_{ij} = {\rm min}(a_i/a_j,a_j/a_i)$, and the angular orbital 
elements of the planets, $\mathbf{\theta}_{ij} = (\lambda_i,e_i,\omega_i,I_i,\Omega_i,\lambda_j,e_j,\omega_j,I_j,\Omega_j)$.  The evaluation of these functions can be found in a series of papers on \hbindex{perturbation theory}: \cite{2008ApJ...688..636N,2009ApJ...701.1116N,2010ApJ...709L..44N,Lithwick2012,2014ApJ...790...58N,Agol2016,Deck2016,2016ApJ...823...72N,Hadden2016,Judkovsky2022a}.

%  - Linear TTV (independently adds from different planets, off resonance)  - 
With the addition of multiple perturbing planets, if the mass-ratios of the planets to the star are
sufficiently small and if none of the pairs of planets are in a mean-motion resonance, then the TTVs may be approximately expressed as linear combinations of the perturbations due to each companion.
For $N$ planets, the TTVs become
\begin{equation}
\delta t_i = P_i \sum_{j \ne i}  \frac{m_j}{m_0} f_{ij}(\alpha_{ij},\mathbf{\theta}_{ij}),
\end{equation}
for $i=1,...,N$.

The largest TTVs are caused by orbital period changes associated with \hbindex{libration} of the system about a mean-motion resonance.  Energy trades can be used to compute the amplitude of the TTV in each planet (see \citealt{Agol2005,2010Sci...330...51H}).  Because of Kepler's relation $a \propto P^{3/2}$, a period lengthening of $\delta P_1 \ll P_1$ is associated with a semi-major axis change of $\delta a_1 = (3/2) a_1 \delta P_1 / P_1$.  Differentiating the orbital energy equation $E_1=-G M m_1 /(2a_1)$ shows that such a change results in an energy change of $\delta E_1=(GMm_1 a_1^{-2}/2) \delta a_1$.  To conserve total energy, the other planet will have an energy change of $\delta E_2=-(GMm_1 a_1^{-2}/2) \delta a_1$, which can also be expressed as $+(GMm_2 a_2^{-2}/2)\delta a_2$.  Using the relation $\delta a_2 = (3/2) a_2 \delta P_2 / P_2$, and the Keplerian relation $a_2/a_1=(P_2/P_1)^{2/3}$, one obtains: 
\begin{equation}
\delta P_2 = -\delta P_1 (m_1/m_2) (P_2/P_1)^{5/3}. \label{eqn:deltaP}
\end{equation}
When considering the $O-C$ shapes that each planet makes over a fixed time interval (e.g. from a survey that measures transits for both planets), there will be a factor of $P_2/P_1$ more orbital periods for the inner planet than the outer planet.  Thus the accumulated time shift of the signal, $\delta t$, builds up more for the inner planet, by one factor of the period ratio. In consideration of equation~\ref{eqn:deltaP}, one is left with: 
\begin{equation}
\delta t_2 = -\delta t_1 (m_1/m_2) (P_2/P_1)^{2/3}. \label{eqn:deltat}
\end{equation}
This scaling agrees with analytic work performed in the resonant \citep{2016ApJ...823...72N} and near-resonant \citep{Lithwick2012,Hadden2016} regimes. Hence the TTV curves of the two planets are \added{often} anti-correlated, with the ratio of planetary masses determining the ratio of TTV amplitudes. In the case that the masses are equal, the amplitude of the outer planet's TTV is larger because its orbital size needs to change more for its \hbindex{Keplerian orbital energy} to equal the change in the inner planet's Keplerian orbital energy. 

In general, transit timing variations afford a means of measuring the density of exoplanets.
The two observables associated with a light curve are the time stamp of each photometric
measurement and the number of photons measured.  The number of photons is a dimensionless
number, and thus may only constrain dimensionless quantities, such as radius ratio, impact 
parameter, or the ratio of the stellar size to the semi-major axis.  The quantities that 
have units of time --- the period, transit duration and ingress duration ---  can 
constrain the density of the system since the \hbindex{dynamical time} relates to stellar density, $\rho$, as
$t_{dyn} \approx (G\rho)^{-1/2}$.  \citet{Seager2003} showed that a single transiting planet
on a well-measured circular orbit may be used to gauge the density of the star;
in the case of multiple transiting planets, the circular assumption may be relaxed
\citep{2014MNRAS.440.2164K}.

The transit depth, then, gives the radius-ratio of the planet to the star, while if two planets
transit and show TTVs, their TTVs give an estimate of the mass ratio of the perturbing planet
to the star.  Thus, two transiting, interacting planets yield an estimate of the density ratio of
the planets to the star, and consequently one obtains the density of the planets.
Note that this is true even if the absolute mass and radius of the star are poorly
constrained.  A caveat to this technique is that there is an eccentricity dependence that 
is present in the stellar density estimate.  However, multi-transiting planet systems typically require low eccentricities to be stable,
and in some cases the eccentricities can be constrained sufficiently from TTV analysis, from
analyzing multiple planets \citep{2014MNRAS.440.2164K}, or
from statistical analysis of an ensemble of planets \citep{Hadden2017}.  So this caveat ends up not impacting the stellar density 
estimate significantly (the mass-eccentricity degeneracy, however, reduces precision on planet-star mass ratios, and hence inflates the planet density uncertainty). 
Another way to obtain an estimate of stellar density is from asteroseismology:
in fact, the time-dependence of asteroseismic measurements is what enables density
to be constrained in that case as well \citep{1986ApJ...306L..37U}.  \added{More recently the Gaia spacecraft has measured accurate parallaxes, which can yield stellar parameters when combined with photometric and spectroscopic data  \citep{Berger2018,Fulton2018}, including density.}

\section{Theory and Paradigmatic Examples} 

This section reviews the physical models for different types of TTV interactions and points the reader to real systems that exhibit each kind of interaction.

Close to resonances, a combination of changes in semi-major axis and eccentricity leads to TTV cycles whose period depends on the separation from the resonance \citep{Steffen2006,Lithwick2012}; the latter refer to this as the \hbindex{`super-period.'}  The \added{dominant} TTV variation comes from only one resonance, the one the system is closest to, which allows its critical angles to move slowly and thus its effect to build up.  If the period ratio $P_2/P_1$ is within a few percent of the ratio $j/k$, with $j$ and $k$ being integers, then the expected TTV period is 
\begin{equation}
P_\mathrm{TTV} = 1/|j/P_2-k/P_1|. \label{eqn:pttv}
\end{equation}
The order of the resonance is $|j-k|$, and the strength of \added{the resonant TTV} depends on the planetary eccentricities to a power of the order minus 1 \added{\citep{Lithwick2012,Agol2016,Deck2016}}.  Therefore, first order resonances affect planets with no initial eccentricity, but higher order resonances have a large effect only in the presence of some eccentricity. 

Seeing two planets transit the star helps immensely to characterize a near-resonant system, because then the relative transit phase of the two planets can be compared with the phase of the TTV signals \citep{Lithwick2012}.  If the eccentricities are maximally damped out, then the resonant terms of the interaction continue forcing a small eccentricity that quickly precesses, causing the TTV.  In this case, the phase of the signal is predictable, and the two planets' eccentricities are anti-aligned, so the TTV signals consist of anti-correlated sinusoids.  Also useful in this case is that the amplitudes lead directly to the planetary masses.  If the so-called \hbindex{free eccentricity} remains, however, the phases would usually differ from that prediction, the TTV in the two planets may not be in perfect anti-phase, and only an approximate mass scale rather than a measurement is available, which is referred to as the \hbindex{mass-eccentricity degeneracy.}  The first real system that showed this pattern convincingly was Kepler-18 \citep{2011ApJS..197....7C}. %[figure of that rather than the theory one given currently in figure 2?]. 
The degeneracy between mass and eccentricity results from sampling at the period of the transiting planet, which causes short period variations to be aliased with $P_\mathrm{TTV}$ \citep{Lithwick2012,Deck2015}.  %\added{\citet{Linial2018} show that TTVs may be decomposed into three components which are comprised of the }

The measurement of TTVs and TDVs has been used for confirmation, detection, and characterization of
transiting exoplanets and their companions.  The \emph{Kepler} spacecraft discovered thousands of transiting
exoplanet candidates;  the classification as ``candidate" was cautiously used to allow for other
possible explanations, such as a blend of a foreground star and a background eclipsing binary causing
an apparent transit-like signal.  The presence of multiple transiting planets around the same star
gave a means of confirming two planets that display {\em anti-correlated} TTVs: due to energy conservation (equation~\ref{eqn:deltat}), the anti-correlation indicates dynamical interactions between the
two planets, while such a configuration would not be stable for a triple star system.  Many papers
used this technique to confirm that \emph{Kepler} planet candidates were bona fide exoplanets using different techniques to identify the anticorrelation in data \citep{2012ApJ...750..113F,2012ApJ...756..185F,2012ApJ...750..114F,Steffen2012a,2013ApJS..208...22X}.

Beyond \hbindex{confirmation} and detection, \hbindex{characterization} of exoplanets with TTVs also began in earnest with the \emph{Kepler} spacecraft.
In addition to Kepler-9, the Kepler-18 system was characterized by a combination of TTVs and
RVs, giving density estimates for the three transiting planets \citep{2011ApJS..197....7C} and assuring that the new method for mass characterization gave the same answers as the trusted, older method.

When only one planet transits in a near-resonant system, the measured TTVs may simply record a sinusoidal signal, which could result from the other planet being close to many different resonances with the transiting planet \citep{2010ApJ...718..543M}.  In Kepler-19, \cite{2011ApJ...743..200B} were able to tell that a planetary companion was the only sensible cause of the TTV, but they were not able to break this finite set of degeneracies\added{; nevertheless, this constituted the first detection of an exoplanet with TTVs. This companion was confirmed, and its period and mass determined, by \cite{2017Malavolta}.} 

This degeneracy has made it extremely difficult to characterize non-transiting planets via TTV, and hence in many cases an additional planet is suspected due to TTV, but detailed work has not been pursued to determine its nature.  The first case of a non-transiting planet being discovered \emph{and} completely characterized was Kepler-46 (a.k.a. KOI-872; \citealt{2012Sci...336.1133N}).  The authors found that the TTVs of the transiting planet were far from a sinusoidal shape; in fact, they could be Fourier-decomposed into at least four significant sinusoids.  Each of these sinusoids can be identified as the interaction with the non-transiting planet via a different resonance.  Even with all this extra information, TTVs could only narrow down the possible perturbing planets to a degenerate set of two, and below is described how TDVs broke this degeneracy.  

Planets that are truly in resonance with each other have the largest TTV signals.  On a medium-baseline timescale like that of \emph{Kepler}, they can perturb each other's orbital periods.  The resonant interaction traps the planets at a specific period ratio, causing the periods to oscillate near that ratio.  The period of the full cycle of that oscillation depends on the ratio of the planet masses to the host star's mass, to the $-\nicefrac{2}{3}$ power \citep{Agol2005,2016ApJ...823...72N}.  For instance, the touchstone system GJ876 has a 550 day libration cycle, about 10 times the outer planet's period, due to its relatively massive planets and low-mass star.  A system which was characterized by resonant interaction is KOI-142 \citep{2013ApJ...777....3N}, in which a non-transiting planet was discovered.   A system with two transiting planets in resonance with large TTVs is Kepler-30 \citep{2012ApJ...750..114F}.  A system with smaller libration amplitudes, but a surprising \emph{four} planets in resonance (forming a chain of resonances) is Kepler-223 \citep{2016Natur.533..509M}. 
\added{\citet{Dawson2021} showed that TOI-216 b/c are in resonance with a joint analysis of the RV and TTVs.}

Several other TTV mechanisms have been detected which do not rely on resonances, but are relevant for more hierarchical situations ($P_2/P_1 \gtrsim 4$). 

If the outer planet transits, and the inner orbiting body is very massive, the dominant effect can be the shifting of the primary star with respect to the \hbindex{barycenter.}   Then, as the outer planet orbits the barycenter, it arrives at the moving target either early or late.  This effect was numbered (i) by \cite{Agol2005}, and it is seen clearly in circumbinary planet systems.  For instance, the secondary star of Kepler-16 \citep{2011Sci...333.1602D} moves the primary by many times its own radius, resulting in an $\sim 8$~day TTV on top of a 225 day orbit. \added{\citet{Millholland2016} have detected a candidate hot-Jupiter, KOI-1822.02, using TTVs induced on an outer transiting companion by reflex motion of the host star, in addition to a phase variation.}

A final mechanism of dynamical TTV is relevant for the inner orbit when a massive body orbits at large distance.  The tide that body exerts on the inner orbit causes its orbital period to differ slightly from what it would be in the absence of that outer body.  If the outer body is in the plane of the inner orbit, its tide slows down the inner orbit, lengthening its period.  If the outer body is far out of the plane of the inner orbit, its tide speeds up the inner orbit, shortening its period.  The tide also depends on the third power of the distance to that external body.  Hence, when the external body moves on an eccentric and/or inclined orbit, it induces a period variation in the inner orbit, which has the period of the outer orbit.  Also important for the timing is how the outer perturber instantaneously torques the inner orbit's eccentricity.  These effects were put together and analyzed by \cite{Borkovits2003} in the context of triple star systems, and the in-plane physics was explained as mechanism (ii) of \cite{Agol2005}.  An example of these effects was provided by Kepler-419 \citep{2014ApJ...791...89D}, in which an eccentric massive planet accompanies an inner planet with a period ratio of 9.7.  \added{A highly-eccentric, long-period ($P \approx 225$ d) transiting Jupiter found with TESS, TOI-4562b, shows TTVS which may be explained by a perturbing companion with a period of 3990 d or nearly 11 years, TOI-4562 c \citep{Heitzmann2023,Fermiano2024}.  If confirmed, this detection may set a record both for the period-ratio of an outer planet found with TTVs ($P_c/P_b \approx 17$), as well as for the absolute orbital period of the companion, $P_c \approx 11$ yr. Such a massive planet
with long orbital period may be detectable with Gaia \citep{Heitzmann2023}, which would enable an estimate of the mutual inclination of these two giant planets.}

\subsection{Chopping}

When two planets are nearly resonant, the degeneracy between the mass ratios of the planets to the star
and the eccentricity vector may be broken by examining additional TTV components present in the data \citep{Deck2015}.
Non-resonant perturbations occur on the time from one conjuction of the planets to the next, which is
when their separation is smallest and gravitational attraction is strongest.  Conjunctions
occur on a period of $P_{\rm syn}= (1/P_1-1/P_2)^{-1}$, also referred to as the synodic period.
TTVs at the synodic period, and its harmonics, have smaller amplitude due to 
the fact that they do not \added{always} add coherently, and thus require higher signal-to-noise to detect.
These synodic variations are referred to as \hbindex{``chopping"} as they commonly
show TTVs that alternate early and late, on top of the larger amplitude TTVs with period $P_\mathrm{TTV}$.
Despite the smaller amplitude, the chopping components can be detected in many cases, and can break
the mass-eccentricity degeneracy, leading to a unique measurement of the masses of the exoplanets 
\citep{2014ApJ...790...58N,2014ApJ...795..167S,Deck2015,Linial2018}.

\begin{figure}
\centerline{
\includegraphics[width=0.9\textwidth]{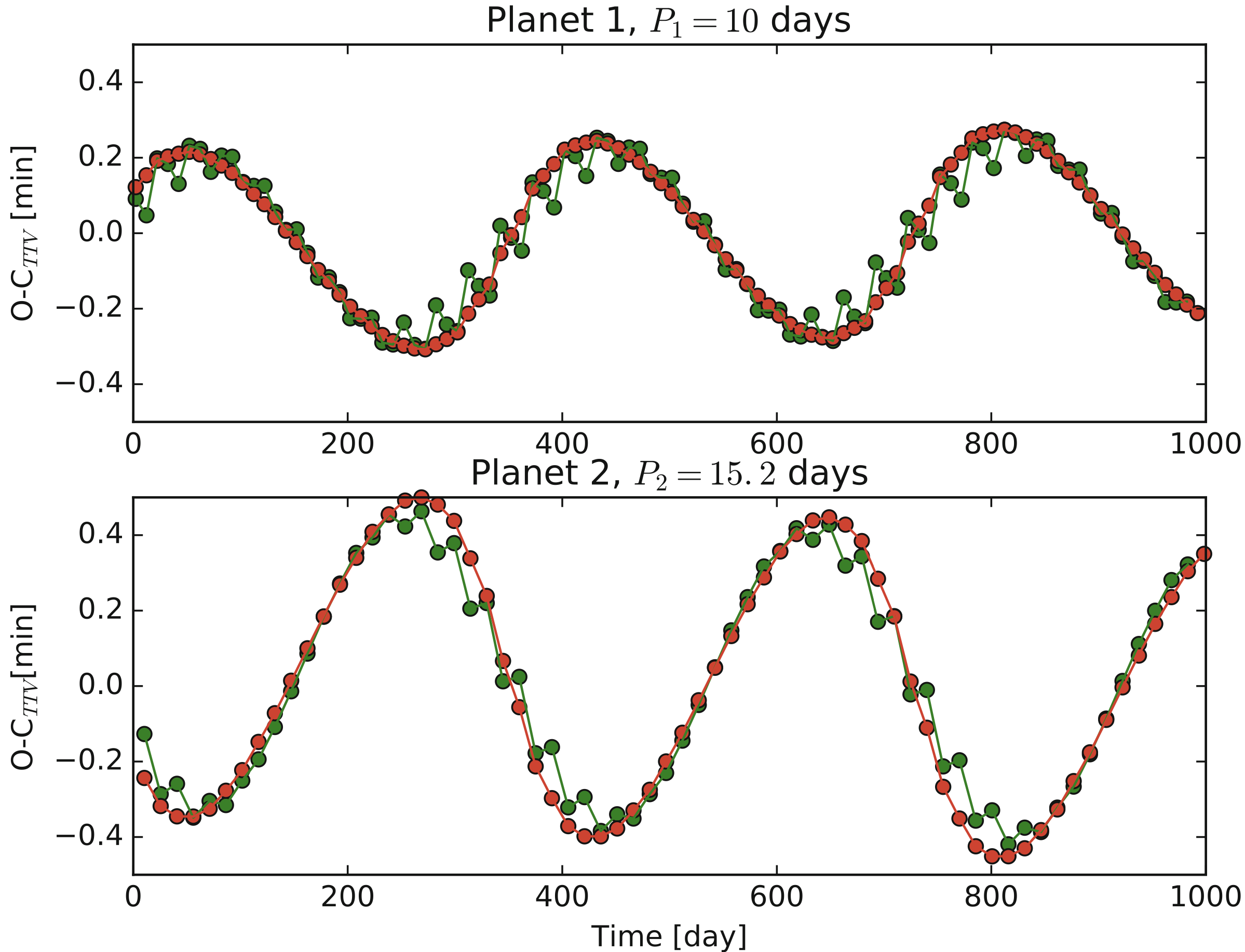}}
\caption{Transit-timing variations of two low-eccentricity planets with larger
mass ratios, $m_1 = m_2 = 10^{-6} m_*$ (green) compared with two higher eccentricity planets ($e_1=e_2=0.04$)
with smaller mass ratios $m_1 = m_2 = 10^{-7} m_*$ (red).  The zig-zag chopping component
is apparent in the high-mass/low-eccentricity case, while less apparent in the low-mass/
high-eccentricity case.}
\label{ttv_chopping}       % Give a unique label
\end{figure}

As an example, consider a pair of planets with period ratio of $P_2/P_1 = 1.52$.  This period ratio
is close to 3:2, and thus is affected by this resonant term, giving a TTV period of $38 P_1$ by equation~\ref{eqn:pttv}.
Figure \ref{ttv_chopping} compares two planets with this period ratio with zero eccentricity
and mass-ratios of $10^{-6}$ to a pair of planets with eccentricites of $e_1=e_2=0.04$
and mass-ratios near $10^{-7}$.  Both pairs of planets give nearly identical amplitudes
for the large resonant term due to the mass-eccentricity degeneracy discussed above, while the larger mass ratio planets show a 
much stronger chopping variation.  In this case there is a clear difference between the TTVs of the two simulated
systems:  the inner planet shows a drift over three orbital periods, and a sudden jump
every third orbital period, while the outer one shows a similar pattern over two orbital
periods.  
In this example the phase of the orbital parameters is set such that the TTV amplitudes match;  change
in the phase can also be indicative of a non-zero eccentricity contributing to the TTVs,
and with an ensemble of planets which are believed to have a similar eccentricity
distribution, the mass-eccentricity degeneracy may be broken statistically \citep{Lithwick2012,
Hadden2014}.

\subsection{Transit Duration Variations}

TDVs have given useful results for characterization of individual systems, though fewer in number than TTVs.  Three mechanisms for TDV have been observed in planetary system orbiting a single primary star.  

The first is torque due to the rotational \hbindex{oblateness} of the star.  It is a convincing model for the duration changes in Kepler-13 b \citep[KOI 13.01][]{Szab2012} and a controversial explanation for transit shape anomalies in PTFO 8-8695 \citep{2013ApJ...774...53B}.  

The second planetary cause of TDVs is eccentricity variations due to a resonant interaction.  The length of the chord across the star, and the speed at which the planet moves along that chord, are changed during the planetary interaction.  This effect has been observed in KOI-142 \citep{2013ApJ...777....3N}.  Slow, \hbindex{secular precession} of the eccentricity is expected by general relativity \citep{2008MNRAS.389..191P}, by stellar oblateness \citep{Heyl2007}, and by tidal distortion \citep{2009ApJ...698.1778R}, but these mechanisms have not given rise to observable TDV to date, for planets around single stars.  It is likely that very long time-baseline measurements, or comparing the measurements of two time-separated space missions like Kepler and Plato, will be able to detect this effect.

The third cause of TDVs for planets around single stars is inclination changes due to secular precession of the orbital plane.  \added{Torques from other planets were observed in Kepler-117 \citep{Almenara2015} and Kepler-108 \citep{Mills2017}, the latter indicating mutual inclination of $\sim 24^\circ$ in a rather hierarchical pair of planets. A more dramatic example is that of KOI-120.01, which went from fully transiting to no eclipses at all during the Kepler mission; \cite{Judkovsky2020} interpret it as a short-period planet perturbed by a few-AU binary companion. Similar phenomena in TOI-216b enabled a better radius measurement to reveal low bulk density for that planet \citep{McKee2023}.}

Earlier the case of Kepler-46 was described, in which TTV measurements of a transiting planet led to two degenerate possibilities for the identity of an additional, non-transiting, planet. The clever resolution \citep{2012Sci...336.1133N} was to note that in one of those solutions, to get the relative amplitudes of the component sinusoids correct in the TTV signal, the perturbing planet must be somewhat inclined with respect to the transiting planet.  As a consequence, a torque on that planet would drive TDV.  No such TDV were observed, so the unique solution --- which is at a different orbital period and planetary mass, and closer to coplanar --- was found. 

Extending this inclination mechanism of TDV to two stars and a planet, the precession of \hbindex{circumbinary planets} (CBPs) has been so extreme as to cause transits to turn on and off \citep{2017MNRAS.465.3235M}.   This observation is similar to the several known cases of stellar triples with inner sometimes-eclipsing binaries, but in this case it is most observable in the outer orbit.  The first case of that phenomenon was Kepler-35 \citep{2012Natur.481..475W}, and the most spectacular observed so far is Kepler-413 \citep{2014ApJ...784...14K}, in which a $4^\circ$ mutual inclination caused transits to stop and then start again nearly half a precession cycle later.  \added{More recently, the first circumbinary planet system with three transiting planets was found with the discovery of Kepler-47d, whose depth grew with time thanks to the changing impact parameter \citep{Orosz2019}.}

Additional dramatic TDVs can occur in CBP systems due to the moving-target effect described above for TTVs.  If the transit occurs while the star is moving in the same direction as the planet, the transit duration is longer; if in opposite directions, the transit duration is shorter.  Matching the prediction from the phase of the binary completely secures the interpretation of the signal that an object is in a circumbinary orbit, as discussed extensively by \cite{2013ApJ...770...52K} for the cases of Kepler-47 and Kepler-64 (a.k.a. PH-1, KIC 4862625b).

\added{As part of the TTV catalog of all the Kepler long-cadence data, \cite{Holczer2016} measured TDVs for the planets with large enough S/N, and \cite{2021Shahaf} described its statistical properties, finding 15 with significant secular trends. They also showed that statistically, the longer-period planets must have more massive or more inclined companions planets in comparison to the closer-in ones. \cite{Millholland2021} also analyzed these data, with conclusions described below. } 

\section{Observational considerations: timing precision} % EA

\begin{figure}
% For example, with the graphicx style use
\centerline{
\includegraphics[width=0.5\textwidth]{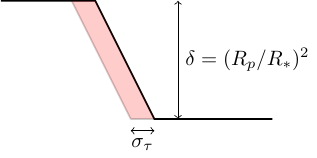}}
\caption{Diagram of the transit ingress of a planet, flux versus time.  The precision of the timing of ingress, $\sigma_\tau$, is set by
when the area of the ingress (pink) equals the timing precision over the duration of ingress. The same applies to egress, albeit
with the time flipped in this plot.}
\label{fig:ingress}       % Give a unique label
\end{figure}

\added{Transit light curves are constructed from a series of exposures, each yielding a flux measurement with an accompanying uncertainty.  The dip in stellar flux during a transit causes a deficit of detected photons relative to the baseline flux outside of transit;  this deficit is fit with a transit model which computes the loss of light caused by the overlapping disks of the planet and star as a function of time.  So, how does this sequence of flux measurements translate into a time of transit?  And how do the uncertainties in flux translate into an uncertainty in the time of transit?  The \hbindex{transit time} is determined from the mid-point of a transit model which best fits the flux time series.  The timing uncertainty is determined by how much the model can be shifted back and forth in time without degrading the fit signficantly.} \added{Thus, the \hbindex{timing precision} is primarily controlled by t}he steepest portions of a transit\added{, which} are the ingress and egress when the planet crosses onto and
off of the disk of the star, causing a dip of depth $\delta = (R_p/R_*)^2$ if limb-darkening is ignored.   
Suppose for the moment that the only source of noise is Poisson noise due to the count rate of 
the star, $\dot N$.  The photometric uncertainty over the duration of ingress, $\tau$ (eqn.\ \ref{ingress}), 
scales as $(\dot N \tau)^{1/2}$.  If the time of 
ingress fit from a model is offset by $\sigma_\tau$, then the difference in counts observed
versus the model is $\sigma_\tau \delta \dot N$ (the pink region in Fig.\ \ref{fig:ingress}).  Equating 
this count deficit to the photometric uncertainty gives \begin{equation}\label{eqn:timing_precision}
\sigma_\tau = \tau^{1/2} \dot N^{-1/2} \delta^{-1},
\end{equation}
which
is the 68.3\% confidence timing precision assuming that the exposure time is much shorter than the
\hbindex{ingress} duration and that $\sigma_\tau \ll \tau$.  The same formula applies to \hbindex{egress.} A longer transit 
ingress duration leads to a shallower slope in ingress, which makes it more difficult to measure an 
offset in time of the model.  Higher count rates and deeper transits improve the precision, as 
expected.  Note that it has been assumed that the duration of the transit is sufficiently long that the error on $\delta$ is small.

Suppose the transit duration is $T$.  Then, the uncertainty on the duration is given by
the sum of the uncertainties on the ingress and egress, added in quadrature:
$\sigma_T = \sqrt{2} \sigma_\tau$.  The timing precision, $\sigma_t$, is set by the mean of the ingress
and egress, giving $\sigma_t = \frac{1}{\sqrt{2}} \sigma_\tau$.

A more complete derivation of these expressions is given by \citet{2008ApJ...689..499C}, while an 
expression which includes the effects of a finite integration time is given by \citet{Price2014}.
The assumptions of no limb-darkening and Poisson noise are generally broken by stars;  in addition,
stellar variability contributes to timing uncertainty, for which there is yet to be a general
expression.  These effects generally increase the uncertainty on the measurement of transit times
and durations, and so the best practice would be to estimate the timing uncertainties from the
data, accounting for effects of correlated stellar variability by including the full covariance
matrix of the timing uncertainty \citep{2009ApJ...704...51C,2012MNRAS.419.2683G,ForemanMackey2017}.  Crossing of
the path of the planet across star spots may also cause some uncertainty on the timing precision 
\citep{2013A&A...556A..19O,Barros2013};
this can be diagnosed by a larger scatter within transit than outside transit or other signs of
significant stellar activity, and can be handled best by including the spots in the transit model 
\citep{2016A&A...585A..72I}.

Note that the barycentric light-travel time offset must be carefully corrected for high-precision
TTV \citep{Fabrycky2010,2010PASP..122..935E}.

\added{In some cases the timing variations can be sufficiently large that the search for transits is impaired as standard algorithms \citep[such as Boxed Least Squares][]{Kovacs2002} assume a linear ephemeris \citep{GarciaMelendo2011}.  An early example of this was the planet Kepler-36b which was strongly perturbed by a close companion.  Each transit was low signal-to-noise, and the Kepler pipeline initially missed this planet thanks to the non-linear ephemeris.  A quasi-periodic pulse detection algorithm (``QATS") was adapted to search for non-linear transits, and identified this planet before the Kepler pipeline later identified the planet with a larger dataset \citep{Carter2013}.  This pair of planets was then precisely characterized thanks to the proximity to 6:7 period-ratio \citep{Carter2012}.  Other algorithms have been developed to search for non-linear ephemerides, most recently using a machine-learning approach \citep[``RIVERS"][]{Leleu2021,Leleu2022,Leleu2023}.}

\section{Science Results}

\added{Initial papers on TTVs were focused on the detection of additional planets in transiting planetary systems.  This was  borne out with the detections of Kepler-19 c and Kepler-46 c, as discussed above \citep{2011ApJ...743..200B,2012Sci...336.1133N}.  However, the number of TTV detections is still modest \citep{Christiansen2022}, with only 32 confirmed planet detections as of October 2024 according to the NASA Exoplanet Archive \citep{NexSci2020}.  A more productive application of TTVs has been towards measuring the masses of multi-transiting planetary systems. }

\added{One of the greatest challenges in the field of exoplanets is the measurement of the masses of Earth-like exoplanets.  This typically relies on measuring the Doppler reflex motion of a star, which becomes prohibitive for an Earth-Sun exo-twin:  the best radial velocity amplitude precisions to date are typically 10-20 cm/sec, while the Sun's radial velocity due to Earth is $\approx$ 9 cm/sec.  Improvements in RV precision have run into a \hbindex{noise floor} caused by stellar variability and inhomogeneity; a path to breaking through this noise floor may be difficult, or impossible, to realize.}

\added{Alternatively, TTVs provide another path towards measuring the masses of Earth-like exoplanets. 
However, there are two drawbacks for using the TTV method to measure planets' masses: 1).  it requires transiting planet(s); 2). it requires the planets to be in close enough proximity that the TTVs exceed the timing noise.}

\added{Both RV and TTV can benefit if one can stretch the definition of ``Earth-like" to include host stars of smaller mass and size:  the smaller stellar mass causes larger reflex motions of the star and timing variations the planets, which increases the amplitude of the RV and TTV signals.  If the meaning of ``Earth-like" also includes locating the planet in the habitable-zone, then low-mass stars have the additional advantage of a closer habitable zone, corresponding to a larger RV signal and a much higher probability of transit.  In addition, a smaller star leads to a larger transit depth, which yields more precise transit times (Eqn.\ \ref{eqn:timing_precision}).
}

\begin{figure}
    \centering
    \includegraphics[width=0.9\textwidth]{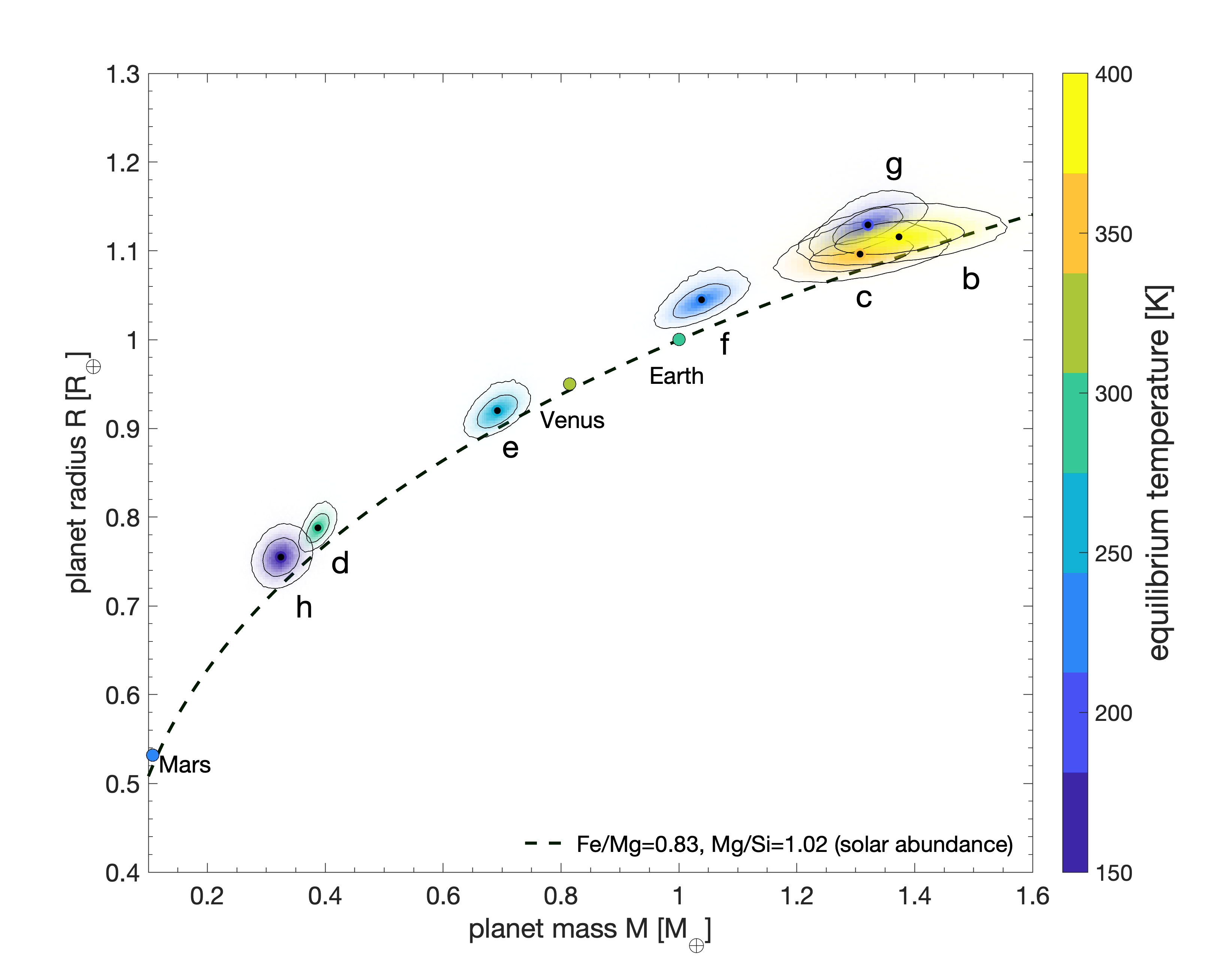}
    \caption{\added{Mass-radius relation of the TRAPPIST-1 planets based on \citet{Agol2021a}.}}
    \label{fig:trappist1_MR}
\end{figure}

\added{Based on these considerations, the discovery of TRAPPIST-1 was a welcome addition to the family of \hbindex{multi-transiting planetary systems} \citep{Gillon2016}.  Eventually seven planets were found to transit the star \citep{Gillon2017,Luger2017}, and the close proximity of their orbits and the small mass and radius of the host star made these promising for characterization of the planets with TTVs over several years of observation \citep{Grimm2018,Agol2021a}.  In fact, these planets' masses can be characterized so precisely with TTVs that the RV technique may never match the precision, and for some of the planets the mass precision is limited by the precision with which one knows the host star's mass.  As discussed above, the transit depths and TTVs give model-independent \hbindex{bulk densities} of the planets (but possibly still corrupted by the inhomogeneous brightness of the star).  These density measurements yield a surprising result:  the planets appear to be consistent with sharing a common composition, but their compositions differ from the Solar System's terrestrial planets (see \citealt{Gillon2024} and
%\underline{TRAPPIST-1 and its compact system of temperate rocky planets} 
Fig.\ \ref{fig:trappist1_MR}).  
If the planets have a smaller ratio of core to mantle mass compared to Earth, then their common compositions might be attributable to a deficit of iron relative to magnesium and silicon of the host star and birth planetary nebula relative to the Sun.  Unfortunately astronomers have yet to measure the abundances of these \hbindex{refractory elements} in the composition of the host star, and this interpretation is not unique.}

\added{Nevertheless, it is remarkable that in the TRAPPIST-1 case one can even address these questions of terrestrial composition thanks to the exquisite precision of the TTVs. For planet e, which is perhaps the most similar to Earth, the measured mass precision would require a 2 millimeter-per-second radial velocity uncertainty to achieve the same precision for a planet at 1 AU around a Solar twin.  This is about two orders of magnitude more precise than the best current RV measurements, which exemplifies to the sensitivity of the transit-timing technique when the right conditions are present. }

A wide-spread phenomenon was detected by TTV characterization of masses: the existence of puffy sub-Neptune planets \added{\hbindex{(``super-puffs")}}.  In the first such case, Kepler-11 e \citep{2011Natur.470...53L}, a planet whose mass is half Neptune's but whose radius is slightly bigger than Neptune's radius.  Even more extreme cases of this class have been found, such as the $2.1^{+1.5}_{-0.8} M_\oplus$ planet with a radius of $7 R_\oplus$ in Kepler-51 b \citep{2014ApJ...783...53M}.  These voluminous envelopes mean these low-mass planets formed while gas was still present in the protoplanetary disk, and that they were able to capture that gas, a surprising result \citep[e.g.][]{2016ApJ...817...90L,2016ApJ...825...29G}.  \added{Alternatively, dusty outflows, high altitude haze, or heating by obliquity tides may be able to explain the large apparent sizes of these planets \citep{Wang2019,Gao2020,Millholland2019}.}

Catalogs of transit times have been produced for the multi-planet \emph{Kepler} systems \citep{Mazeh2013,Rowe2015,Holczer2016,Ofir2018,Thompson2018,Lissauer2024}. Several analyses of an ensemble of TTV pairs of planets have recently been carried out \citep{Hadden2014,2013ApJS..208...22X,2014ApJS..210...25X,JontofHutter2016,Judkovsky2022b,Judkovsky2024,Ofir2025}, with the largest by \citet{Hadden2016}, %Hadden \& Lithwick (2016), 
yielding constraints on the RMS eccentricity of the population of planets. At periods near $\approx 10$ days, the RV and TTV densities agree rather well.  At shorter period, most of the RV detections are single planets, which in general appear to have lower density relative to their multi-planet counterparts \citep{Steffen2016,2017ApJ...839L...8M}.

With the end of the primary \emph{Kepler} mission, the data volume of transit times diminished \added{until the launch of \hbindex{TESS}}. \added{And yet analyses of Kepler data continue to provide new results. \citet{Millholland2021} used TDVs to argue that the so-called \hbindex{``Kepler-dichotomy",} which is the observation from transiting planetary systems that the geometrical bias applied to statistical models of the multiple-planet systems falls short of explaining all the single-transiting planets \citep{2012Johansen}, arguing for a model of two populations \citep{Ballard2016},}\added{may instead be explained by a single population in which there is a anti-correlation between multiplicity and mutual inclination of planets; this is also confirmed with TTV analyses of the population, indicating that 30\% of stars have planetary systems like those found with Kepler \citep{Zhu2018}. \citet{Yee2021} show that the Kepler systems analyzed with TTVs show eccentricities which are smaller than they would be if the systems were randomly drawn from the stable phase-space, indicating that some sort of damping occurs; also see \citet{VanEylen2015} and \citet{VanEylen2019}.  \citet{Judkovsky2022b} show that photodynamical models can detect variations in the impact-parameter of Kepler planets, which manifest as TDVs, caused by companion planets or other effects. \citet{Ofir2025} applied an analytic model to analyze more than 200 planets from the Kepler mission with photodynamics.  A search for TTVs of hot jupiters in the Kepler data reveal that some of these systems may have close companions \citep{Wu2023}. In addition, various systems have been subject to joint reanalysis with additional transit data from the full Kepler dataset, and/or combined analyses with follow-up RV and/or transit observations  \citep{vonEssen2018, Freudenthal2018, SaadOlivera2017, SaadOlivera2018, Sun2019, Freudenthal2019,Jackson2019,Masuda2020,Petit2020, Vissapragada2020,Weiss2020,Liang2021, MacDonald2021,MacDonald2022,Sun2022, GreklekMcKeon2023}.  The Kepler dataset is a gift that keeps on giving for timing analyses.} %(e.g.\ lots of examples - Kepler-9, Kepler-46 Saad-Olivera et al., Kepler-90, KOI-984 etc.)

\added{After completion of the Kepler mission,} the \hbindex{K2 mission} continued to provide TTV systems, such as WASP-47, the first
short-period hot Jupiter with nearby planet companions \citep{Becker2015},
\added{as well as multi-planet systems K2-138 \citep{Vivien2024}, K2-146 \citep{Hamann2019}, and K2-24 \citep{Nascimbeni2024,Teyssandier2020}.  With the end of the K2 mission and launch of TESS in 2018} \citep{Ricker2015}, \added{astronomers are presently in a new era of studying transit-timing with the brightest stars on the sky which are also amenable to radial-velocity follow-up.  These observations are being accompanied by novel theoretical developments, described next.} 

\section{\added{Recent Theoretical Developments}}

\added{Modeling of dynamically interacting planetary systems has continued to progress since the first version of this review.  This includes algorithms for the detection of TTVs, characterization of systems with TTVs/TDVs, and interpretation of the dynamics of known dynamically-interacting systems.}

\added{\citet{Ofir2018} developed a spectral approach to detecting TTVs for pairs of interacting planets, which they then applied to the Kepler dataset, resulting in a significant increase in the number of TTV detections.}

\added{As mentioned above, \citet{Leleu2021} developed a new machine-learning approach to detecting transiting planets which display large TTVs, and applied it to Kepler data to detect new planets \citep{Leleu2021,Leleu2022}.  This method also reveals a possible bias when carrying out transit-timing analyses due to under-estimating the TTVs \citep{Leleu2023}.}

\added{Bayesian sampling of the dynamical parameters of transiting planetary systems can be improved with faster N-body algorithms.  A \hbindex{symplectic} approach was taken by \citet[][``TTVFast"]{Deck2014}, which was given a Python interface by \citet[][``nauyaca"]{Canul2021}.  Other numerical codes for analyzing TTVs have been developed, such as \citet[][``TRADES"]{Borsato2014} based on a Runge-Kutta integrator.}

\added{An open-source code based on perturbation-theory was provided by  \citet[][``TTVFaster"]{Agol2016}.  An approximate semi-analytic approach was taken by \citet{Linial2018} who decompose the TTVs of a pair of planets into three ``modes," after which a linear fit can yield the planets' masses, based on the insights of \citet{Lithwick2012}.  \citet{Hadden2019b} developed an analytic model for computing TTVs, ``TTV2Fast2Furious," while \citet{Hadden2022} give an example of computing TTVs with their Python code for perturbation theory, ``celmech."}

\added{\hbindex{Photodynamical} models enable more precise characterization of planets with shallow transits, such as Kepler-36 \citep{Carter2012}, and can yield more precise masses and densities.   An analytic photodynamical model was developed and applied by \citet{Yoffe2021} and \citet{Judkovsky2022a,Judkovsky2022b, Judkovsky2024}, yielding more precise masses, radii, and densities \citep{Judkovsky2023}. In systems with extreme TDVs, including disappearing transits, a fast analytic model has been developed to characterize the dynamics by \citet{Judkovsky2020}.  These analytic models can be useful, but they are always approximate, and will become inaccurate in certain parts of parameter space, at which point one needs to rely on N-body computation. }

\added{In systems with many planets, the high dimensionality of the model parameter space can prohibit standard \hbindex{markov-chain monte carlo} techniques from converging quickly to the posterior probability.  In addition, the optimization of models can be inhibited by inaccurate finite-difference derivatives, which are also slow to compute \citep{Rein2016}.  Thus, the accurate computation of derivatives of transit-times with respect to initial conditions can greatly improve the computation of the maximum likelihood and likelihood profile, and improve posterior probability sampling using Hamiltonian/Hybrid Monte Carlo (HMC) \citep{Duane1987} and/or the No U-turn Sampler \citep[][``NUTS"]{Hoffman2014}.  The first transit-timing code to implement derivatives analytically was developed in the  Julia language for speed and ease of use \citep[][``NbodyGradient.jl"]{Agol2021b}.  This code made possible the sampling of the TRAPPIST-1 system in a small amount of CPU time \citep{Agol2021a}.  More recently, Kento Masuda has developed a symplectic transit-timing model based on \hbindex{automatic differentiation} \citep{Baydin2018}  in the JAX language \citep[][``jnkepler"]{Dai2023,Jones2024,Masuda2024}.  Differentiable photodynamical models have also been developed as part of jnkepler (in JAX) and in Julia \citep[][``Photodynamics.jl"]{Langford2024}.  The JAX \citep{Jax2018} and Julia \citep{Bezanson2017} languages provide user interfaces which are interactive and high-level, but also use just-in-time compiling to provide code which runs fast, solving the ``two-language" problem.  These codes will help to improve statistical analyses of the parameters of dynamically-interacting planetary systems.}

\added{Once a system is analyzed, its dynamical state can be characterized.  \citet{MacDonald2023} developed a code ``exoMMR" for diagnosing the possible presence of mean-motion resonances, and they also forecast the expected TTV amplitudes for systems with detected resonances. Several papers have examined the statistics of period-ratios and TTV ``phase" in pairs of Kepler planets \citep{Goldberg2023,Choksi2023,Wu2024,Li2024}.  These papers propose that the TTV phase, caused by non-zero free eccentricity, may be excited by a variety of mechanisms:  secular forcing by more distant companions, tubulence in the protoplanetary disk, scattering by small bodies, or giant impacts.  None of these theories yet provides a perfect match to the data, but some make predictions that may be tested with further examination of the dynamical states of these systems.}

\added{\citet{Delisle2017} has shown that TTVs may be used to probe planetary spin dynamics, while \citet{Boley2020} showed that the shortest timescale eigenfrequency can determine the rate of change of TDVs, which can cause some transits to (dis-)appear with time.}

\added{Extreme TTVs/TDVs can occur for binary planets in which two planets orbit about their center of mass, which in turn orbits about the star.  This leads to large timing and duration variations, as modelled by \citet{Chakraborty2022} and \citet{Martin2019}.}

\added{These theoretical developments will help to address a new era of TTV/TDV analysis being ushered in by TESS, described next.}

\section{\added{Transit-timing in the TESS era}}

\added{The TESS mission is currently in operation, with each sector covered for about one sidereal lunar month at a time, while the polar caps are covered for up to one year of contiguous sectors \citep{Ricker2014}.  Despite the short duration of each sector relative to Kepler, this mission has discovered planets transiting some of the brightest stars on the sky, making these stars amenable to transit-timing follow-up with ground-based telescopes and other space-based telescopes, such as CHEOPS \citep{2014SPIE.9143E..2JF,Borsato2021b}. Despite its smaller aperture, TESS has been able to detect transits for some Kepler planet systems, extending the time-baseline significantly \citep{JontofHutter2021,Battley2021,JontofHutter2022,Yahalomi2023}. The longer duration of observations at the polar caps and the multi-year coverage as TESS alternates between the Southern and Northern hemispheres enables the longer-term monitoring necessary for TTV analyses, which was anticipated by \citet{Hadden2019a}.
This has made TESS a prolific telescope for the discovery and characterization of systems displaying TTVs. Here are highlighted a number recent discoveries.}

\added{Examples of pairs of transiting planets which are in mean-motion resonance are rare, yet TESS provided an early example with the pair of planets TOI-216b/c \citep{Kipping2019,Dawson2019,Dawson2021,Nesvorn2022}.  The inner planet was initially grazing, preventing a measurement of the planet's radius, but with further observations the planet is now fully transiting the disk of the star, reveal the planet's radius, which implies a low density;  this is an extreme example of TDVs \citep{McKee2023}.
A non-transiting companion planet was detected from large transit-timing variations of transiting planet TOI-2015b \citep{Jones2024}, but as with Kepler-19c, the period of the perturbing planet is ambiguous.  %The period of the perturbing planet is ambiguous, despite the significant amplitude of the TTVs (100+ minutes) being detected at high significance, with candidate periods close to the 4:3, 3:2, or 2:1 period-ratios with the transiting planet.  TESS detected the transiting planet in multiple sectors, while follow-up was obtained from a range of ground-based telescopes. Further transit-timing and/or more precise radial velocity measurements may be able to break the current transit-timing degeneracy in this system.  
Another planet detected with TTVs by TESS is TOI-199c \citep{Hobson2023}, which may reside in its star's \hbindex{habitable zone.}  Other systems display TTVs for larger transiting planets, including TOI-2525 \citep{Trifonov2023}, TOI-2202 \citep{Trifonov2021,Rice2023}, TOI-1803 \citep{Zingales2025}, and TOI-5404b \citep{Vitkova2025}; the latter planet shows the largest TTVs of any planet to date with variations of $\pm$2 days!}

\added{\hbindex{Resonant chains} are rare, but treasured for helping to constrain formation and migration processes in high-multiplicity planetary systems.  TESS found several 6-planet systems with some or all of the planets in a resonant chain:  TOI-1136 is a system of six transiting exoplanets \citep{Dai2023}; TOI-178 also contains six planets, of which the outer five appear to be in a resonant chain \citep{Leleu2021,Delrez2023,Leleu2024a}; and HD 110067 \citep{Luque2023} has all six planets in a resonant chain \citep{Lammers2024}.  }

\added{Other TESS systems do not reside in resonant chains, but do display TTVs which have been employed to constrain the planet masses and orbits, such as TOI-270 \citep{Kaye2021}, TOI-1246 \citep{Turtelboom2022}, and TOI-1130 \citep{Huang2020,Borsato2024}.  The system TOI-1130 consists of a hot Jupiter (``c") with an interior Neptune-mass planet (``b"), both with percent precision on their masses and radii; a transit of planet b was missed in follow-up observations, indicating the presence of large TTVs induced by the outer companion, confirmed by later observations which showed large TTVs, enabling a photodynamical characterization of the system \citep{Korth2023,Borsato2024}.  Two of the planets in the TOI-2076 system show anti-correlated TTVs, but which do not yet enable a precise mass constraint \citep{Osborn2022}. The planet TOI-2818b shows TTVs, but does not yet have sufficient data to determine the nature of the companion, while RV upper limits rule out some TTV solutions \citep{McKee2025}. TOI-1749 shows hints of TTVs in the outer two planets \citep{Fukui2021}, without sufficient data to make a clear TTV characterization.  Most recently TTVs have been detected for planets transiting a star which is visible to the human eye \citep{Bonfanti2025}.  Some systems are expected to show detectable TTVs, but lack sufficient observations, such as TOI-125 \citep{Quinn2019}, TOI-561 \citep{Lacedelli2020, Weiss2021}, TIC 279401253 \citep{Bozhilov2023},  TOI-2096 \citep{Pozuelos2023}, HD 152843 \citep{Nicholson2024}, and TOI-700 \citep{Gilbert2020,Gilbert2023}.  The latter system is remarkable in that it hosts two Earth-sized planets close to a 4:3 period ratio in the host M-dwarf star's habitable-zone. The system LP 791-18 has a host star which is bright enough for follow-up transit-timing observations with ground-based telescopes, which yield a mass of the temperate Earth-sized planet in the system and constrains its \hbindex{tidal heating} rate \citep{Peterson2023}.  In the case of TOI-1266, \citet{Cloutier2023} used the mass measurements from RV and the \emph{lack} of TTV to place a more stringent upper limit on the eccentricities of the two inner planets than the limit placed by RV.}

\added{With its all-sky coverage, TESS has provided access to younger transiting planetary systems, such as AU Microscopi, which contains an edge-on \hbindex{circumstellar disk} and has been followed up with transit-timing measurements using CHEOPS \citep{Plavchan2020,Wittrock2022, Wittrock2023,Martioli2021, Szab2021,Szab2022}.  HIP 41378 f shows a long-term quadratic trend in the transit times, which likely needs more data to derive a dynamic solution \citep{Bryant2021,Alam2022}.  The young planet TOI-1227 b shows a long-term timing trend \citep{Almenara2024}.  %A remarkable transiting planet with a period of 225-days and eccentricity of 0.76 was found around the star TOI-4562 \citep{Heitzmann2023}.  The transit-timing variations of planet b have been interpreted as a massive giant planet with an orbital period of nearly 11 years \citep{Fermiano2024}, which would be a record for any planet detected with TTVs.   Such a massive planet with long orbital period may be detectable with Gaia \citep{Heitzmann2023}, which would enable an estimate of the mutual inclination of these two giant planets. 
Each of these systems offers the prospect of characterization of the masses of these planets with both radial-velocity and transit-timing, which has a significant advantage of being able to check that both techniques agree, instead of being confused by stellar variability or other systematics (see below).}

\added{What can one learn from these systems?  TESS has given stronger evidence that multi-planet systems have evolved dynamically over time.  Younger systems appear to have a higher incidence of pristine resonances \citep{Dai2024}, while the period ratios of older systems can be modeled with a mix of resonances and broken resonances caused by planetary collisions \citep{Li2024}.  Since the sizes of planets evolve with time, either due to cooling or envelope loss, the masses of planets are needed to diagnose the evolution over time, which TTV can provide in the case of closely-spaced multi-transiting planetary systems.  The early evolution of systems due to \hbindex{planetary migration} can even be constrained with the architecture of these systems.  In TOI-1136, the proximity of two of these planets to a 7:5 period ratio informs the system's migration history due to the delicate nature of second-order resonances \citep{Dai2023}.}

\added{Early work on transit-timing carried out searches for companions of hot-Jupiters, many of which turned up empty-handed \citep{Steffen2012b}; TESS seems to confirm this \citep{Zhang2024}.  Gradually companions to hot Jupiters were found with Kepler and K2 \citep{Wu2023}, and now the TESS spacecraft has found its first examples of interior companions to hot-Jupiters, TOI-1130b/c (mentioned above) and TOI-1408b/c \citep{Korth2024}.  The Jupiter-sized planet TOI-1408b shows grazing transits with a period of 4.4 days, which was found by the standard TESS pipeline.  The second planet was missed by the TESS pipeline, and later found to transit with custom pipelines, due to its much smaller size and due to the largest fractional TTVs of any system found to date.  The presence of nearby companions to hot Jupiters places strong constraints on models of their formation and migration \citep{Dawson2018}.}

%\added{TESS has yet to find significant examples of transiting planets with significant TDVs, and yet their bright host stars may allow for detections with follow-up observations.}

\section{\added{Transit-timing versus radial velocity}}

\added{The transit-timing and radial-velocity techniques can both suffer from systematic errors which can make the inferred masses incorrect.  The most common errors are due to additional planets or stellar activity noise, but other issues can affect either technique, such as complications with the instrument, data-incompleteness, analysis, and/or modeling inacccuracies.  Thus, a cross-comparison between these techniques can help to bolster our confidence in both techniques.  Such a comparison has been limited, but may be about to grow significantly with the advent of the TESS mission.}

\added{With the Kepler mission came the first detections and growth of the transit-timing technique for measuring masses of exoplanets.  However, the Kepler field covers a small part of the sky with few stars which allow for measurement of their planetary masses with radial velocity. Radial velocity surveys instead target the brightest stars across the sky to achieve smaller photon-noise, as well as to better characterize the stellar jitter using spectral diagnostics.  Thus, in the Kepler era there were only a handful of planets with both transit-timing mass measurements {\it and} radial velocity measurements, allowing for limited comparison of the two techniques.}

\added{Sample-level comparisons between RV and TTVs circumvented the incompatibility of Kepler and RV surveys \citep[e.g.][]{Wolfgang2012,Zeng2019}, but \hbindex{selection effects} can plague these comparisons \citep{Steffen2016,2017ApJ...839L...8M,JontofHutter2019,Otegi2020,Zhu2020}, and overcoming these biases requires careful correction \citep{Leleu2023,Leleu2024b}.  Since the TESS spacecraft is now covering the brightest stars on the entire sky, transiting-planet hosts can now be characterized with radial-velocity measurements (as well as with astrometry and direct-imaging).  Since transits have a low probability, transiting planet hosts are typically fainter than host stars of radial velocity surveys.  However, with the advent of the ESPRESSO instrument on the VLT \citep{Pepe2021}, high precision RV is now possible with many transiting planet host stars. The CARMENES instrument on Calar Alto 3.5-meter observatory  \citep{Quirrenbach2014} is also extending high-precision RV measurements to M dwarf stars, which are also fainter, but are better represented in the TESS sample than in the Kepler sample.}

\begin{figure}
    \centering
    \includegraphics[width=\linewidth]{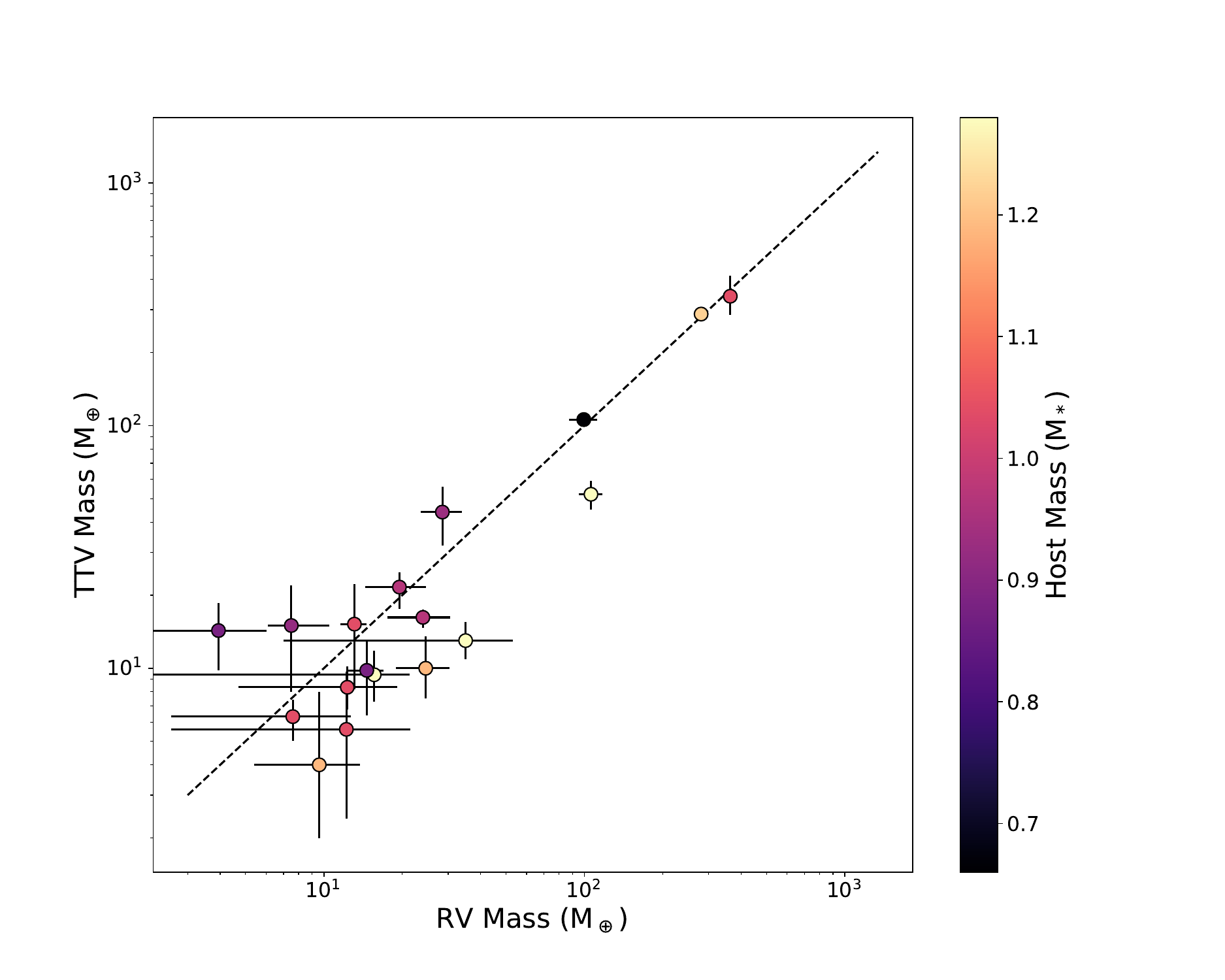}
    \caption{\added{Plot of measured masses of transiting exoplanets with TTV and RV measurements.  (Figure credit: Tyler Gordon)}}
    \label{fig:RV_vs_TTV}
\end{figure}

\added{Direct comparisons of TTV and RV mass-measurments are now available for a growing number of systems.  Figure \ref{fig:RV_vs_TTV} shows a selection of planets taken from the literature for which the masses have been measured separately with TTVs {\it and} RV.  There is a good correlation between the two methods; in some cases one technique gives better precision than the other, but in general the masses are consistent, which is good news for both techniques.  One outlier is Kepler-89d/KOI 94.01 \citep{Weiss2013,Masuda2013} in which the TTV mass is smaller by factor of $\approx 2$ relative to the RV mass.  This is system with four transiting planets, for which the best-fit TTV model is a poor fit.  \citet{JontofHutter2022} find a slightly larger mass, reducing the tension of the fit, but more work remains to determine if, perhaps, adding an additional planet to the system can produce a better TTV fit, agree with the RV data, and resolve the tension between the RV and TTV masses.}

\begin{figure}
    \centering
    \includegraphics[width=\linewidth]{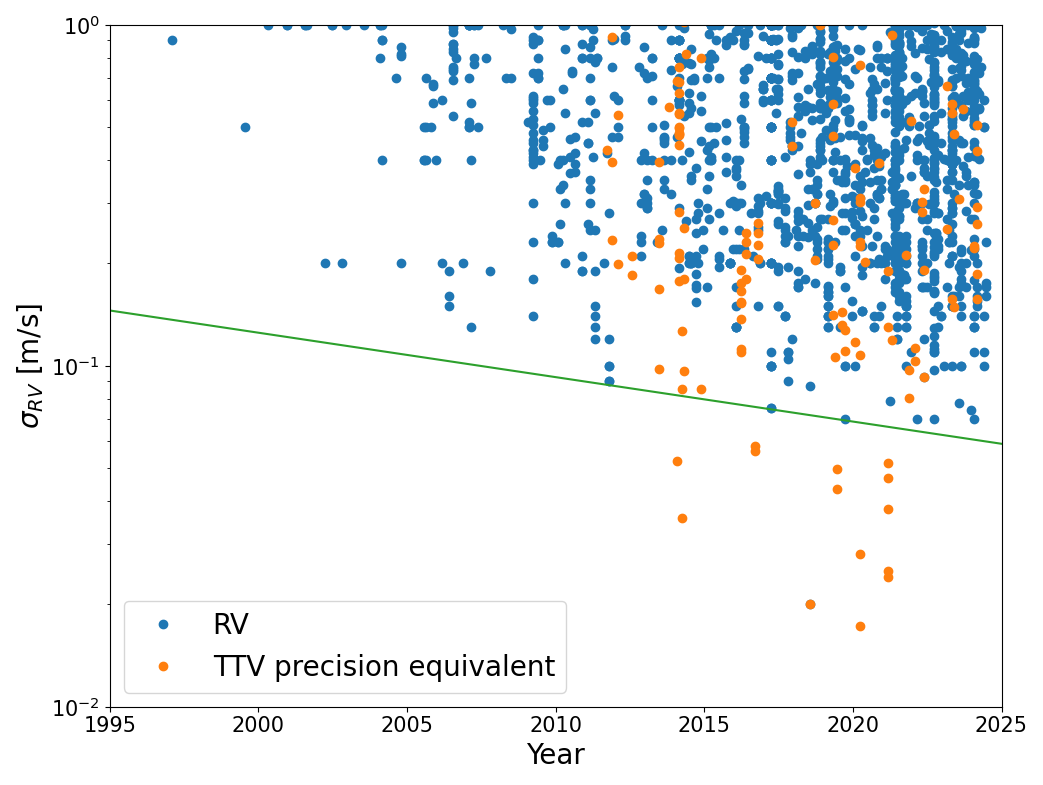}
    \caption{\added{Radial-velocity semi-amplitude precision (blue) and semi-amplitude-equivalent for TTV mass measurements (orange).  Green line shows the improvement in RV precision with time.  All of the highest precision measurements below 6 cm/sec belong to the TTV technique.}}
    \label{fig:RV_TTV_precision}
\end{figure}

\added{One way to contrast transit-timing with radial-velocity is to compute the equivalent radial-velocity semi-amplitude uncertainty required to obtain the same mass-uncertainty as a transit-timing measurement.  Based on data from the NASA exoplanet archive, the RV-equivalent precision was computed for TTV-measured masses (some of these TTV masses may be limited by the measurement precision of the host star), as well as the RV-semi-amplitude precision for planets with masses measured with RV.  After correcting several typos in the database and in published papers, Figure \ref{fig:RV_TTV_precision} shows that radial-velocity measurements have been improving versus year of publication.  However, the highest equivalent precision of the two techniques is given by TTVs.  A drawback for precision TTV mass measurements is that they require extensive observations, so no measurements better than 6 cm/sec equivalent have been published since 2021.}

\added{This figure may indicate that TTV could be used in the future as a calibrator of extreme-precision radial velocity (ePRV) measurements.  As larger telescopes with better instrumentation continue to improve RV precision, the effects of stellar variability can dominate the noise, and detrending or mitigating these effects can be complicated.  If TTV systems found around bright stars can be characterized with high precision, perhaps in the future these systems may be used to calibrate ePRV detrending techniques.}

\section{\added{Transit-timing with HST and JWST}}

%\added{Follow-up of transiting planets with Hubble and JWST }
%\added{Exomoon candidate show TTVs which could instead be due to another planet (Kipping, Yahalomi, etc.).  }

\added{Much of the exoplanet focus of HST and JWST has been trained on the detection and characterization of exoplanet atmospheres thanks to their photometric stability, near-infrared spectroscopic capability, and, for JWST, mid-infrared sensitivity \citep{2014PASP..126.1134B}.  Despite the power of HST and JWST, for many planets the detection of an atmosphere cannot be accomplished in a single visit, but must build up signal-to-noise with multiple transit and/or secondary eclipse measurements.  For example, a significant portion of JWST time has been devoted to characterizing the atmospheres of the TRAPPIST-1 system planets with observations of multiple transits and secondary eclipses.  The multiple visits to each of the planets in this system will provide transit times and durations, which will likely yield more precise constraints on the dynamical properties of the system than with the Spitzer telescope thanks to the much larger collecting area and sensitivity to the peak of the star's SED.  The secondary eclipses can provide a complementary constraint on the orbital eccentricities of the planets \citep{Winn2010,Heyl2007}, which can potentially break TTV degeneracies.}

\added{Serendipity can also allow HST or JWST to detect TTVs.  The HST transit of Kepler-1625b showed a significant timing offset relative to the Kepler ephemeris \citep{Teachey2018}, likely due to a planetary companion, or, more intriguingly, potentially due to a candidate exomoon. The transits of the planets Kepler-51 b and d \citep{LibbyRoberts2020,Masuda2024} show timing offsets with HST and JWST. A timing analysis implies that a fourth planet must be present in the system, while more data will be required to disentangle the nature of this system of low-density planets.}

\added{Another aspect of transit-timing which can be probed with JWST is wavelength-dependent times of transit.  This  effect was predicted by \citet{DobbsDixon2012} for simulated hot-Jupiter transit-transmission spectra.  The offsets are not due to the dynamics of the orbital motion, but instead are due to atmospheric dynamics.  Stellar irradiation can drive a super-rotating circumplanetary jet, which causes the Eastern terminator to be hotter than the Western, causing a difference in scale-heights along the two sides of the planet.  The wavelength-dependence of the opacity then causes the shadow of the planet to appear distorted and offset at some wavelengths, causing an artificial timing offset when the transits are fit with a circular planet shadow.  This effect was detected for the first time with JWST by \citet{Rustamkulov2023} in the hot Jupiter WASP-39b, which displayed wavelength-dependent timing offsets which were qualitatively consistent with the predictions of \citet{DobbsDixon2012}.}

\section{\added{A prescription for transit-timing/duration analyses}}

\added{Here is a brief guide for transit-timing and duration analyses. The first step involves measuring the times and durations of the individual transits. Usually this requires fitting a transit light curve to the transits, often holding the radius-ratio fixed, and allowing the transit time, duration, and impact parameter to vary, and then sampling these parameters from the posterior probability distribution.  Many tools exist to carry out these analyses, as described in other chapters, such as EXOFASTv2, exoplanet, and jaxoplanet \citep{Eastman2019,ForemanMackey2021,Hattori2024}.}

\added{The next critical component of the analysis involves determining if there are detectable TTVs/TDVs.  This requires confidence in the error bars, and then seeing whether the times (durations) are consistent with a linear ephemeris (constant value).  If the chi-square is large for these, then it could be indicative of the presence of TTVs/TDVs; it could also mean that the error bars are underestimated, or that there are outliers present. Usually a large number of measurements are necessary for redundancy in checking for outliers;  in addition, a sufficiently long sampling timescale is needed for an accurate analysis.  It is best to have a duration of two super-periods which is well-sampled.  If two planets are in close proximity and are close to a resonance, then the TTV amplitude can be estimated based on mass-radius relations and analytic formulae.}

\added{If one is confident of the detection of TTVs/TDVs, and would like to characterize the system, then next a model for the timing must be developed.  Analytic approaches are much faster computationally, and can be useful for an initial analysis, but they are approximate,  so a full analysis should also be carried out with N-body integration.  This usually involves integrating Newton's force laws with time, and then finding the minimum projected sky separation to determine the mid-transit time \citep{Fabrycky2010}.  An example of computing TTVs is given in the open-source REBOUND N-body code github repository \citep{Rein2012}.  An initial model must be found for the TTVs/TDVs, which is often a tricky prospect.  If the period and/or amplitude of the TTVs/TDVs is revealed by fourier analysis, then analytic estimates may be used to seed the dynamical model.  Often this means placing a perturber near a resonance to produce large sinusoidal TTVs (although if more than one planet transits, and companion planets are in close proximity, then this can make this step straightforward).}

\added{Next, the N-body model must be optimized.  This typically involves computing a gradient of the likelihood with respect to the dynamical parameters, and often finite-differences are used for this purpose, although gradient-based computations have become available more recently, either analytic or with automatic differentiation (as discussed above).  The Levenberg-Marquardt algorithm works well when the uncertainties are Normal, while the Broyden-Fletcher-Goldfarb-Shanno algorithm (BFGS) tends to work well otherwise \citep{Fletcher1987}.  Other approaches involve gradient-free optimization, such as Nelder-Mead, or simply running a markov chain and picking the highest likelihood, but these tend to be slower.  With the optimized model, then one must ask again whether the model matches the data.  Frequently the chi-square is large (compared with the number of degrees of freedom), and so again there exists several possibilities:  perhaps the optimum likelihood was not found; perhaps the error bars are underestimated; perhaps the dynamical model is inadequate (often this means another perturbing planet, which can be exciting); or perhaps there are outliers in the data.  The residuals can be examined, both temporally to look for correlated patterns which may indicate an inadequate model, as well as examining the normalized residuals to see whether they look Gaussian, or whether there may be outliers present.  If correlations are not present, and the errors appear normal, but the standard deviation is too large, then often times rescaling of the uncertainties or adding a systematic noise error in quadrature to the measurement errors can suffice to account for a large reduced chi-square.  Once the model is re-optimized with the scaled uncertainties, then the uncertainties on the model parameters can be estimated from the information matrix (note that this step assumes that the likelihood is a multi-dimensional Gaussian, which is often not the case, but it does supply a starting estimate for parameter errors).  Although this has yet to be widely applied to timing analyses, a \hbindex{cross-validation} approach is valuable if there is sufficient data, in which some data are excluded, the model is optimized to the remaining data, and then the model is used to forecast the excluded data.  This approach can be used to test how robust the model solution is, and can possibly be used in flagging outliers, or determining whether to introduce more complexity to the model.  Prior to sampling the full probability, one should carry out a profile of the likelihood.  This involves stepping through each parameter, holding it fixed, and optimizing the likelihood with respect to the remaining parameters.  This can help to identify degeneracies in the model, or help to find other modes of the likelihood (i.e.\ high-probability regions) which may not be sampled well by MCMC.  The widths of the profile of each parameter may be compared with the information matrix error estimates to see whether these are comparable.}

\added{The penultimate step involves sampling the \hbindex{posterior probability} of the model given the data.  The sampling step is only meaningful if the physical model and the statistical model represent the data well, and so care must be taken to address these issues.   Techniques that are standard now are using markov-chain Monte Carlo or nested sampling;  however, these methods can have difficulty converging in high-dimensional parameter spaces.  Again, gradient-based algorithms can enhance convergence with HMC or NUTS.  If the initial optimization revealed multiple solutions with similarly good fits, then sampling each of these ``modes" is necessary before reporting uncertainties.}

\added{Finally, once the posterior parameters are derived, then a dynamical analysis can take place.  Often this involves examining the long-term stability of the system, measuring the free and forced eccentricities, looking for signs of resonance, etc.  If an ensemble of systems are being studied, then a statistical examination of their properties can take place.}

\section{\added{Unsolved problems and future developments}}

\added{Future work on TTVs and TDVs can continue to improve our application of these techniques, and to better understand their planetary systems. One unsolved problem is the presence of excess timing noise in transit-timing measurements.  In some cases the source(s) of this noise have not yet been identified, and may be due to stellar variations (e.g.\ variability of the total flux due to granulation, spot crossings, and/or flares), or potentially due to systematics (either due to instrumental or atmospheric instability).  This excess noise is apparent when comparing the excess timing scatter in single- and multi-transiting planetary systems in the Kepler dataset \citep{Siegel2022}. Robust statistics may be used to handle outliers, such as the \hbindex{Student's-t distribution} \citep{JontofHutter2016,Agol2021a}. Recent JWST observations of TRAPPIST-1 show an excess of low-amplitude stellar flares \citep{Howard2023}, which were below the threshold of detection in prior Spitzer and Kepler data, which may be a culprit for timing outliers.}

\added{Other future work could address theoretical modeling of transit timing.  Analytic models for TTVs may underestimate the ``chopping" component \citep{Deck2015}.  Searching for companion planets requires a global search through parameter space, which can be difficult thanks to the narrow high-probability regions occupied by transit-timing solutions in a high-dimensional parameter space.  Several works have pointed out possible biases in TTV analyses, which is a problem which has yet to be fully explored.  % Analytic transit-timing models are available near resonance, but have yet to be developed for planets which are in resonance, while recent progress has been made on modeling resonant systems \citep{Hadden2019c}.  
While analyzing transit-timing variations, it is typical to consider long-term stability of the system after a posterior has been derived from the timing model \citep[e.g.][ applied to Kepler-36]{Deck2012}.  Incorporating stability criteria directly into the timing analysis could help improve the dynamical characterization \citep[e.g.][]{Tamayo2020}.  Finally, distinguishing TTVs of companion planets from exomoons could use more theoretical investigation to improve ongoing searches for exomoons \citep{Kipping2020a,Kipping2020b,Kipping2022,Yahalomi2024a}.  Machine learning and artificial intelligence has yet to be widely applied to transit-timing and duration variations.  \citet{Chen2024} has developed the ``DeepTTV" model, while   \citet{Ikhsan2024} have investigated the application of machine learning for the TTV inverse problem.  }

\added{Continuing to search for and catalog transit-timing and duration variations is ongoing work.  Kepler catalogs continue to be refined \citep{Lissauer2024}. There has yet to be a catalog for TESS, with the exception of \citet{Ivshina2022} which only examines hot-Jupiters, and finds that only WASP-12 shows evidence of period decay.  Most transit-timing studies are bespoke for each high-multiplicity system, as the analyses are detailed and tedious, and each system seems to contain a surprise or a puzzle.  Streamlining, automating, and producing an assembly line for these analyses would be a valuable endeavor --- perhaps a job for the next generation of AI?}

%\added{Plato, HWO, Nautilus}

\added{This is especially important work as the collection of transit measurements will continue to grow throughout this decade and the next.  The ARIEL space telescope will primarily be focused on spectroscopic characterization of transiting extrasolar planets \citep{Tinetti2018}.  It will greatly improve on CHEOPS thanks to its larger collecting area and its more placid orbit near L2 \citep{Tinetti2016}. It is due to launch in five years time, and will operate for four years, allowing sufficient time for multiple visits of some transiting planets to improve on the signal-to-noise, but also to simultaneously search for TTVs \citep{Borsato2021a}.  The Roman mission (aka WFIRST/WFIRST-AFTA) will not be targeted, but it will revisit the same fields numerous times, with many transiting planet detections expected \citep{2017PASP..129d4401M}, enabling a search for TTVs and TDVs \citep{Wang2023}.}

\added{The PLATO mission is due to launch in 2026 \citep{2014ExA....38..249R}, while the `Earth 2.0' mission is due to launch in 2028 \citep{Ge2022}; these may well cause another spike in TTV science.
  These missions have the prospect of detecting new planets in systems already characterized with Kepler and TESS, as well as extending the timing baseline significantly \citep{Eschen2024}.  On a longer timescale, the Habitable Worlds Observatory may expand on the capabilities of JWST for precision transit spectroscopy \citep{Gaudi2020}, sometimes requiring multiple visits which can then be used for timing analyses.  A concept being studied for a very large aperture set of space telescopes could greatly improve precision spectroscopy of transiting planet systems \citep[][``Nautilus"]{Apai2019}.  This would help in overcoming the limits of stellar variability on transit characterization \citep{Gordon2020}, potentially enabling the future study of solar-system analogues \citep{Lindor2024}.}  

\added{The future of characterizing multi-transiting planet
systems with TTVs and TDVs looks promising.}

\begin{comment}
\section{Cross-References}

\begin{itemize}
\item Transit Photometry as an Exoplanet Discovery Method
\item The Way to Circumbinary Planets
\item Detecting and Characterizing Exomoons and Exorings
\item Space Missions for Exoplanet Science: Kepler/K2
\item Space Missions for Exoplanet Science: TESS
\item Tools for Transit and Radial Velocity Modelling and Analysis
\item Dynamical Evolution of Planetary Systems
\item Tightly Packed Planetary Systems
\item Two Suns in the Sky: The Kepler and Tess Circumbinary Planets
\item TRAPPIST-1 and its compact system of temperate rocky planets
\end{itemize}
\end{comment}

\begin{acknowledgement}
%EA acknowledges support from NASA Grants NNX13AF20G, NNX13A124G, NNX13AF62G, from National Science Foundation (NSF) grant AST-1615315, and from NASA Astrobiology Institute's Virtual Planetary Laboratory, supported by NASA under cooperative agreement NNH05ZDA001C.  
E.A. acknowledges support from NSF grant AST-1907342, NASA NExSS grant No.\ 80NSSC18K0829, and NASA XRP grant 80NSSC21K1111.  This work was completed while D.C.F. was supported by (while serving at) National Science Foundation.  Thanks to Dr.\ Tyler Gordon for compiling the data and creating Figure 5.
\end{acknowledgement}

%  IF you do NOT use bibtex, put comments before the following 2 lines
\bibliographystyle{spbasicHBexo}  %for bibtex
\bibliography{agol_fabrycky} %for bibtex-example

\end{document}